\documentclass[10pt]{article}
\usepackage{PRIMEarxiv}
\usepackage{xcolor}
\usepackage{natbib}
\usepackage{ifthen}
\usepackage{graphicx}
\usepackage{fancyhdr}
\setcitestyle{} 

\newcommand\defaultfont{}

\newcommand\card[1]{{%
\small\raisebox{-0.6ex}{\hspace{0.1em}%
\includegraphics[width=0.8em]{#1.png}
}%
}\hspace{0.1em}\ignorespaces%
}

\newcommand\PAO[1]{\emph{#1}}

\title{Representational Tenets for Memory Athletics}

\newcommand\affilAFRL{\textsuperscript{1}}
\newcommand\affilParallax{\textsuperscript{2}}
\newcommand\affilDylan{\textsuperscript{3}}
\newcommand\affilRazvan{\textsuperscript{4}}
\newcommand\affilMobius{\textsuperscript{5}}

\author{
Kevin Schmidt\affilAFRL\textsuperscript{*},
Othalia Larue\affilParallax,
Ray Kulhanek\affilParallax,
Dylan Flaute\affilDylan,
Razvan Veliche\affilRazvan,
\\\bfseries
Christian Manasseh\affilMobius,
Nelson Dellis,
Scott Clouse\affilAFRL,
Jared Culbertson\affilAFRL,
Steve Rogers\affilAFRL
\\[1.5ex]
\affilAFRL{}Air Force Research Laboratory\hspace{1.0em}
\affilParallax{}Parallax Advanced Research\hspace{1.0em}
\affilDylan{}Jacobs\hspace{1.0em}\\
\affilRazvan{}Keystone Strategy\hspace{1.0em}
\affilMobius{}Mobius Logic
\phantom{0pt}\\
\textsuperscript{*}kevin.schmidt.15@us.af.mil\\
}

\begin{document}
\defaultfont
\maketitle
\thispagestyle{fancy}

\begin{abstract}
We describe the current state of world-class memory competitions, including the methods
used to prepare for and compete in memory competitions, based on the subjective report of World
Memory Championship Grandmaster and co-author Nelson Dellis. We then explore the reported
experiences through the lens of the Simulated, Situated, and Structurally coherent Qualia (S3Q)
theory of consciousness, in order to propose a set of experiments to help further understand the
boundaries of expert memory performance.
\end{abstract}

\section{Introduction and Background}
Analyzing expert cognitive performance on tasks that involve resource-limited constraints allows
researchers to explore the boundaries of computational cognitive models. This approach dates back at
least to the classic works of Herbert Simon and Allen Newell, who advanced theories of Artificial
Intelligence (AI) through symbolic information processing models of expertise
\citep{chase1973perception,newell1981mechanisms}. Newer models of expertise (as summarized in
\citet{gobet2015chunks}) have provided additional fidelity and richer explanations, but settings
such as memory athletics still provide opportunities for refining these models and potentially
guiding novel AI representations.

The purpose of this note is twofold. First, this work describes the current state of
world-class memory competitions, based primarily on the subjective report from World Memory
Championship Grandmaster and co-author Nelson Dellis (ND). We discuss the principal method used to
prepare for and compete in competitions, the Method of Loci, as well as the current neuroscientific
projection of this method onto the multiple memory systems in the brain.  Secondly, this manuscript
will explore the phenomenological aspects of this extreme memory experience through the lens of the
Simulated, Situated, and Structurally coherent Qualia (S3Q) representational framework of
consciousness \citep{schmidt2021like} and show how this framework suggests novel experiments to
further understand the boundaries of expert memory performance. This paper is written to be
accessible to those in cognitive neuroscience who are interested in the particular memory mechanisms
described, but also to those in AI who are interested in implementing related systems.
Previous work has demonstrated the value of syntheses of neuroscience and AI. For example, artificial neural networks have a tendency to forget previously learned information when learning new tasks (catastrophic forgetting). The multiple memory systems within the brain help to limit the disruption of existing structures when learning new information \citep{mcclelland1995there}; understanding these systems might therefore inspire new artificial neural structures that are more resilient to catastrophic forgetting.

\subsection{Speed Cards}
The operating characteristics of these multiple memory systems can be described by assessing elite
memory athletes competing in memory competitions. ``Speed Cards'' is one of the most common events
in memory competitions, and will be used as an example throughout this paper. Competitors are given
up to five minutes to memorize a shuffled standard 52-card deck. They are then given a new deck in
factory order that must be placed in the same sequence as the first deck.  If at least one
competitor reproduces the deck perfectly, the winner is the competitor who took the least time to
memorize the deck (reconstruction time is not included). If all decks contain errors, a judge
compares the original and reconstructed deck one card at a time, and the score is determined by the
number of correct cards before the first mistake \citep{worldmemoryhandbook2013}. Speed Cards world
record holder Shijir-Erdene Bat-Enkh has accomplished a perfect deck in 12.74 seconds. ND, who
achieved a grandmaster ranking at the 2012 World Memory Championships \citep{wmscstats2022},
achieved a time of 40.65 seconds, offering a comparable case study for understanding how such memory
achievements are possible.

\subsection{Expertise and Multiple Memory Systems}
Successful performance in this task is thought to require a complex interaction between the two
types of long-term memory in the human brain. The first depends on a localized biological structure
called the hippocampus, which is needed to consciously experience past episodes from our lives
\citep{squire1992declarative,squire2015conscious}. This episodic representation can be investigated
through the Simulated, Situated, and Structurally coherent Qualia (S3Q) representational framework
of consciousness, which describes qualia as the fundamental units composing an internally generated
spatiotemporal and causal sensorimotor world model (simulation) defined by relationships with other
qualia (situatedness) and grounded through awareness of the environment (structural coherence)
\citep{schmidt2021like}. The central tenet of the S3Q model is a prioritization of stability,
consistency, and usefulness of the representation over fidelity with the environment; the functional
utility of this being the regularization capability of constructing a world model while constraining
representational complexity with mechanisms of parsimony. The S3Q framework provides a way to
express and explore the experiential aspects and underlying mechanisms of conscious memory by
deconstructing the phenomenal experience and characterizing the foundational representation.

In contrast to the characteristics of the conscious qualia-based memory representation, there are
other forms of memory in the brain which are evolutionarily older and reflect a principle of
neuroplasticity that operates nonconsciously, whereby the long term memory network comes to reflect
the statistical regularities of the environment \citep{reber2013neural,squire1992declarative}.
Through repeated activation, the landscape of the neural network is adapted to the regularities of
experience to allow more efficient processing, such as that seen in memory athletes, and much like
current machine learning tools. Nonconscious statistical learning takes place automatically whereas
conscious learning is highly regularized \citep{schneider1977controlled}. The allocation dynamics of
conscious processing resources in the brain are well characterized by multiple resource theory
\citep{wickens2008multiple}, but a deeper understanding is needed of how these limited conscious
resources interact with implicit, nonconscious representations to achieve optimal performance during
memory competitions.

Both of these systems learn in parallel, with the conscious learning system being important for
one-shot learning of object locations, while the mature implicit neural circuits allow for rapid
memory retrieval and fluid pattern-matching decision making
\citep{chase1973perception,klein2010rapid}. For example, chess masters are exceptionally good at
recognizing patterns of meaningful positions of pieces on a chess board and can instantly store in
memory entire chess board arrangements \citep{chase1973perception}. In chess, a sufficiently regular
state-space exists such that regular patterns of play have come to be reflected in experts' long
term memory through chunked retrieval structures, allowing them to instantly match large knowledge
structures in long-term memory to the patterns on the board \citep{gobet1998expert}. A pattern
matching process thus allows the honed neural representation of the expert chess player to rapidly
connect the positions on the board to an advantageous play, similar to the performance seen in
expert radiological diagnosis \citep{kundel1975interpreting,lesgold1988expertise,evans2013gist}.
Through practice, the sets and repetitions have tuned the neural network in the brains of experts
such that statistical patterns in the arrangement of the pieces on the chess board are quickly
recognized as known patterns in the expert's brain \citep{chase1973perception}, allowing them to
efficiently respond appropriately. These pattern-matching feats of chess masters are resilient to
interruptions, signifying that they are drawing on long-term memory for their performance gains, and
not short-term working memory, which is known to be affected by interruptions
\citep{charness1976memory}. These same processes and effects are seen with memory athletes in card
memorization, as will be described below. 

\subsection{Cognitive Models of Expertise}
Earlier models of expertise emphasized short-term memory limits, proposing chunking models of
learning. These theories estimated the limits of this working memory store at seven, or more
recently, four chunks \citep{miller1956magical,cowan2001magical}. Many of these theories explicitly
declared that all learning is acquired through limited capacity short-term memory chunks
\citep{chase1973perception}. Specifically, earlier theories of expertise made the assumption that
there was only one learning system: initially memories are encoded consciously and these
representations are later transformed into automatic processes \citep{anderson1982acquisition}.
However, this approach is not able to account for the rapid acquisition of long-term memory seen in
expert workers. For example, a case study of an expert restaurant waiter documented the ability to
remember the spatial layout of a table of customers (always broken down into tables of four
customers maximum), and then chunk the information of their orders spatially by the customer's seat
at the table, then by food order (e.g., meat, starches), etc. \citep{ericsson1988cognitive}. The
expert waiter is still only able to hold four chunks of information in mind at a time, yet the
structure of that information is nested hierarchically, which is thought to allow retrieval
structures to access and organize many bits of information directly into long-term memory, all while
still being nested within the limited chunking mechanism. 

Without a direct route into long-term memory (i.e., without two systems), earlier chunking theories
could not account for the waiter being able to remember many orders over an extended time amid
distractions. Later models \citep[e.g.,][]{maddox2004dissociating,ashby2005human} offer a parallel,
implicit way into the brain which can be described through mechanisms called \emph{templates}
\citep{ericsson1995long} by which pointers in working memory and retrieval structures in long-term
memory are used to interface with chunks of related information. Thus, specific \emph{slots} are
created that hold typed variables that can be rapidly written to during learning by experts in their
area of expertise (see the example of Person-Action-Object templates below). While the working
memory pointers are thought to be represented (in part) in the hippocampus, the structure of the
information chunks is thought to be represented through statistical mechanisms in the neocortex
unless in interaction with the medial temporal lobe learning system. Thus, there are parallel
representations of learned information: one can be developed nonconsciously and is distributed
pervasively throughout the brain through mechanisms such as Hebbian statistical neuroplasticity,
whereas the other requires the localized medial temporal lobe system to interface with this
underlying neuronal structure to allow for conscious access of the learned content, through
mechanisms such as pattern completion inferencing.

EPAM (Elementary Perceiver and Memoriser \citep{feigenbaum1984epam}) and CHREST (Chunk Hierarchy and
REtrieval Structures \citep{gobet2002search}) are instantiations and implementations of Template
Theory in which  chunks are computationally defined as nodes in a discrimination network that act as
an index to long-term memory. Influenced by the earlier EPAM model, an early theory and cognitive
architecture created by Herbert Simon, modeling how humans learn to recognise patterns, and how
learning affects recall and categorization behavior, CHREST originally modeled chess expertise.
Nodes (chunks) in the discrimination network act as an index to long-term memory. This
discrimination network, a branching tree structure,  is modeled as a hierarchical sequence of
perceptual tests which classifies the features and leads to an existing leaf node in the long-term
memory network which returns the system's knowledge in connection to the stimulus.  During
classification, if a difference is identified in the current discrimination network, information is
added to the node (if there is familiarity) or a new node is created under the previous parent node.
Short-term memory contains references to information held in long-term memory. Frequently occurring
patterns in the environment allow for chunks to evolve into complex data structures, called templates:
a slotted schema that can contain both variable and fixed information.  Templates support expertise
since they can hold variables for the rapid acquisition and integration of new information into the
long-term memory network.

\section{Subjective Report}
Having provided the neuroscientific basis for current understanding of memory management in the
brain, in this section a subjective report is given primarily by ND, capturing experiences in both
memory training and participation in memory competitions. The 
Method of Loci\footnote{This technique is known by several other names, including Memory Palace, Mind Palace, Roman Room
Method, Journey Method, and the Peg Method. While variations occur in the way that the method is
presented and taught, the fundamental concept of linking a mental image with a spatial location is
common across all formats.}
technique is a mnemonic method in which a practitioner visualizes traversing sites in a spatial
location (a ``Memory Palace''), and the items to be remembered are ``placed'' at the sites in that
location \citep{yates1966art}. More specifically for Speed Cards, many memory athletes use a
mnemonic system involving using a Person-Action-Object (PAO) triple encoding to identify three cards
with one experienced image. The athletes then link these images via an ordered memory palace to
store and retrieve those images in a desired sequence. This is the technique used by Speed Cards
world record holder Shijir-Erdene Bat-Enkh \citep{delle2017interview}, as well as by ND. 

\subsection{Summarizing the Strategy}
Specific mnemonic systems have been developed for abstract memorization tasks and there are several
such systems for memorizing cards within the broad umbrella of the Method of Loci. When applied to
Speed Cards, the common goal across these systems is to turn each card in a 52 card deck into
something memorable. Former memory champion Dominic O'Brien developed one such system called the
Dominic System \citep{obrien2013pass}. 

\begin{figure}[t]
\centering
\includegraphics[width=\textwidth]{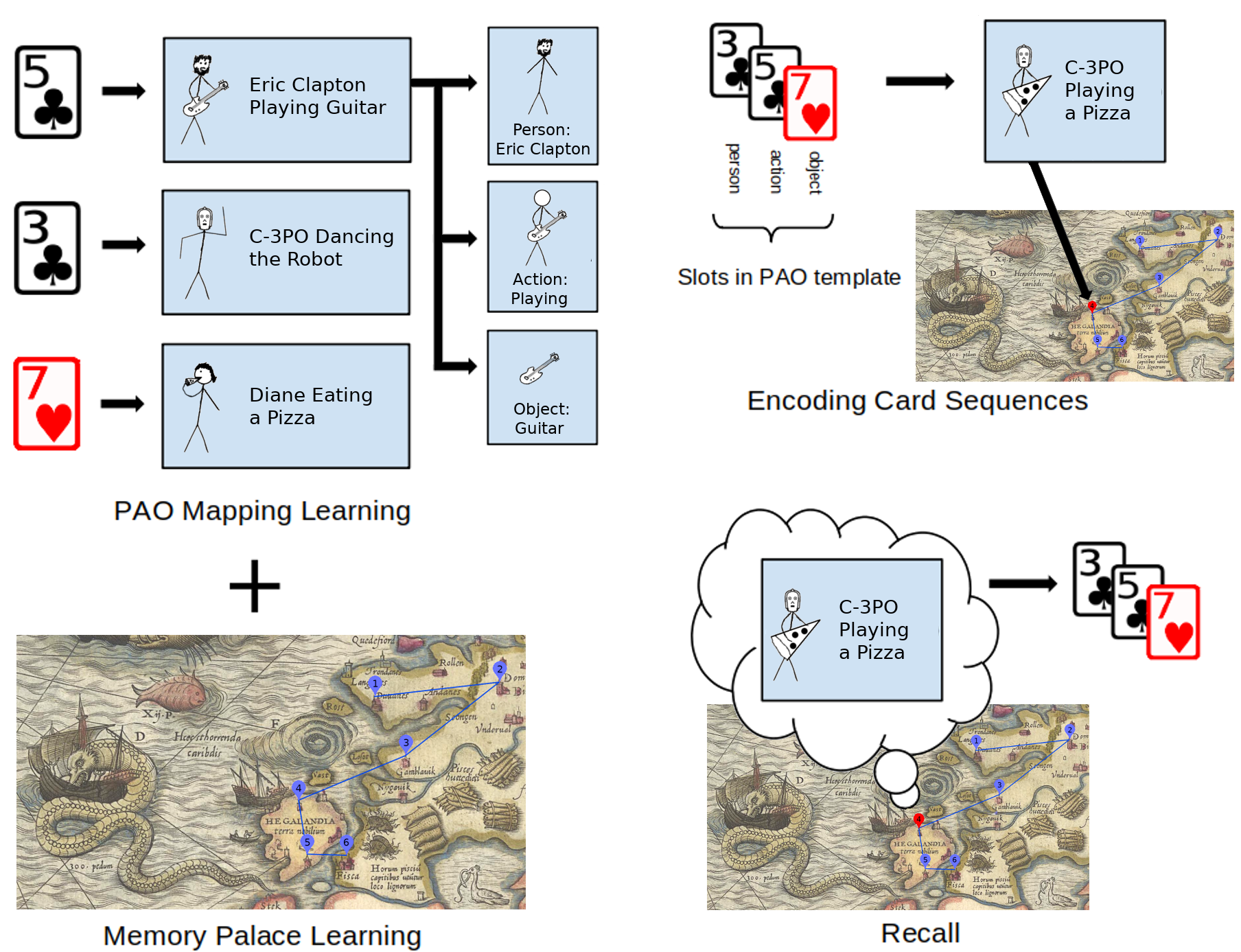}
\caption{Person-Action-Object Method of Loci: 
Before the competition, the memory athlete memorizes a mapping from each card onto the corresponding person, action, and object (top-left)
and a sequence of locations within the memory palace (bottom-left).
During competition, the memory athlete memorizes the deck by mapping each triple of cards onto the corresponding person, action, and object (in that order), and placing the
resulting scene in the next location of the memory palace (top-right). Then during recall, they traverse the memory palace in order, observe the memorized scene at each location, and map it back
to the cards (lower-right).
\citep[Map excerpt:][]{cartamarina}
}
\label{fig:pao}
\end{figure}

The method consists of two components. The first is a \emph{memory palace}: a
pre-memorized familiar location containing several sub-locations (at least 17,
for the Dominic System). For example, ND uses his high school, where the
sub-locations are specific locations throughout the school (e.g., the entrance
to the school, specific classrooms, and the gymnasium). The memory athlete
visualizes themselves visiting each sub-location in a fixed order (ND starts at
the entrance to the school, then proceeds in sequence to certain classrooms)
and observing a scene placed there (Figure \ref{fig:pao}, bottom-left).  

The scenes themselves are constructed
using the Person-Action-Object (PAO) mapping.  Each card is associated with an
arbitrary, but distinctive person.  For each person, an action and an object
that relate to the person are then chosen. For example, in ND's mapping (Figure \ref{fig:pao}, top-left), the five of
clubs could correspond to:
\begin{itemize}
\item \PAO{Eric Clapton} (person)
\item \PAO{playing} an instrument (action)
\item a \PAO{guitar} (object)
\end{itemize}
Both the memory palace and PAO mapping are
fixed, and will be memorized before the competition.

When presented with a deck to memorize at the competition, cards are memorized
in sequences of three by taking the person associated with the first card, the
action associated with the second card, and the object associated with the
third card. This visualization of the person doing the action to or with the
object is then embedded in the next  location of the memory palace. For example, ND
maps the sequence \card{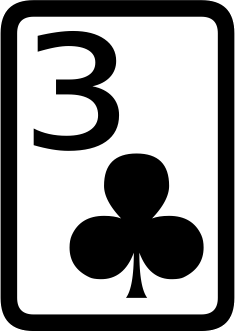}\card{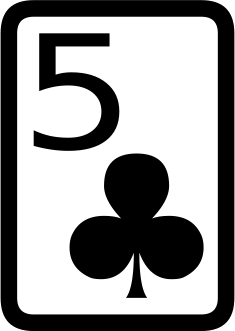}\card{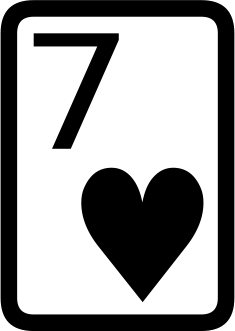}
onto \PAO{C-3PO}
(\card{3C} person) \PAO{playing} (\card{5C} action) a \PAO{pizza} (\card{7H}
object) (Figure \ref{fig:pao}, top-right). During recall, the athlete visits each location in turn, observes the
scene, and recovers the card sequence that it encodes (Figure \ref{fig:pao}, bottom-right). The problem of recalling
52 cards is thus reduced to recalling 17 unique scenes.  (Further detail is
provided in \ref{appendix:dominic}).

\subsection{Distinctiveness of Representation}
Memory athletes may try to reduce the difficulty of discriminating the items in
their memory palaces by ensuring the PAO associations are sufficiently
distinct. When a memory athlete has two people that could be associated with
the same action, the memory athlete may try to choose a different action for
each person that would be a stronger association. For example, ND uses Kobe
Bryant and Shaquille O'Neal, both basketball players, in his PAO scheme. While
Kobe Bryant and Shaquille O'Neal could both be shooting a basketball, ND
instead associates Bryant with dribbling and O'Neal with dunking. ND reports
that this aids in reducing the memorization time for the PAO mapping.

\subsection{Qualitative Experiences During Training and Testing}
During the memorization phase, ND is aware of looking at the three cards
one-by-one rather than looking at all three cards at once, but perceives a
single composite image associated with the three cards. ND compares it to
reading a sentence: he reports that, when encoding a triple of cards in a
certain location, he does not think about the previous three cards or next
three cards at all. In a competition, after the memorization phase and stopping
his timer, ND will use the remaining time to review the memory palace. He
reports that this review is crucial. He traverses the memory palace from the
most recent spot to the first spot (i.e., backwards from the order he memorizes
the cards). He finds that, if he can go through the palace once, he can think
about something else (or even have a conversation) for the rest of the time and
still recall the information. 

ND reports that he spent time making imagery vivid when he was a novice, but,
given the time pressure of competition settings, the amount of detail in the
imagery must drop. This happened automatically, not intentionally, for him. He
gives the example of his father: ND used to take time to visualize his father's
hair, hear his accent, and figure his height based on the location, but now, ND
mainly imagines the ``essence'' of his father's hair. He reports that the parts
he does not focus on are blurry, fuzzy, and grayscale while the things he does
focus on have some color. An interesting example of this is the Ace of Spades.
The person for this card is Arnold Schwarzenegger. ND reports that ND does not
see Schwarzenegger at all and only hears Schwarzenegger's voice saying a
particular thing. When creating the PAO image, the action is performed by a
``blob.''

\section{S3Q-Guided Analysis and Hypotheses}
In order to push the boundaries of our understanding of memory athletes and
memory systems in general, we give both a characterization of the neural
mechanisms used in the Method of Loci as well as a methodology for using
current cognitive modeling tools to explore these mechanisms. Finally, we
propose two hypotheses that highlight the potential for how the S3Q
theory of consciousness could provide meaningful guidance toward novel
understanding of expert memory performance.

\subsection{Neural Characterization}
We hypothesize that high repetitions practice will lead to a decrease in the
network activation strength needed to maintain performance over time, while the
conscious learning mechanism will be indexed by a positive activation signal
for successful memory.  In popular computational models of expertise
\citep{gobet2000five}, it takes 250 ms to write a chunk into a template slot.
\citet{fellner2016spatial}  hypothesize that the memory palace is represented
via $\sim$250 ms theta oscillations of place cell activity in a hippocampal template.
Higher frequency alpha, beta, and gamma ($\sim$10-50 Hz) rhythms could represent the
imagined items translated from the physical cards and chunked into an event.
It is hypothesized that the PAO chunk is either placed into working memory representation via these
theta rhythms, or if templates have been developed, directly into a slot in
long term memory represented by higher frequency rhythms. During competition,
synchronous high frequency oscillations in the posterior cortex are here
hypothesized to ride theta oscillations directly into slots generated by the
hippocampus and surrounding cortex, binding chunks of PAO items to the spatial
context map loci. Neuronal sequences generated by the hippocampus are
hypothesized to act as working memory pointers that interface with retrieval
structures represented in cortical neural networks that have been honed through
extensive practice. Repetitive experience automatically activates Hebbian
neuroplasticity mechanisms (``neurons that fire together, wire together'') to
optimize activity based on statistical demands of the memory event
\citep{hebb2005organization}, creating templates in the hippocampus and
neocortex that are hypothesized to achieve efficiency through statistical
prediction that optimizes the trajectory of the oscillation towards improved
performance. After repeated practice mentally navigating a memory palace, a
template is formed such that forward prediction of sequential loci via pattern
completion facilitates transition between event boundaries
\citep{dubrow2013influence}.

Similarly, a PAO template allows efficient transition 
between variable categories represented in distant cortical regions for the oscillations, chunking the imagined qualitative features matched 
to sensory cue input between variable categories represented in distant cortical regions. 
These processing
efficiencies allow the integration of more features into the same limited
bandwidth theta cycle. The implication is that the frequency bandwidth of the
4-chunk conscious, working memory-based learning mechanism is the same for
experts and novices ($\sim$250 ms), but within their domain of expertise, memory
athletes are better able to access hierarchical long term memory chunks through
efficient retrieval structures honing slots into nested oscillations by high
repetition neuroplasticity learning mechanisms. This trainable capacity can be operationalized
through an increase in theta-gamma cross-frequency coupling during the memory
competition, for example, or a decrease in high frequency power and amplitude as
a function of practice. These parallel mechanisms allow athletes to achieve
increased qualitative capacity with practice, through an interaction of S3Q
resource allocation honed for efficiency by the statistical demands of the
task.

\subsection{Cognitive Modeling for Memory Experimentation}
As described in the introduction, there are distinct learning systems enabling
expert performance during competitive memory athletics. The first is a
conscious representation that flexibly balances S3Q resource tenets for the
Method of Loci during the deck of cards competition. In contrast, a parallel
nonconscious system emerges through practice to adapt neural circuits for
efficient processing at the cost of flexibility. These two learning systems
interact for elite memory performance, providing a clear demonstration that
performance is not fixed, but matures with practice to provide a more rich
conscious experience.

Although other cognitive architectures (SOAR, ICARUS) could be used for this
model, we leveraged some of our previous modeling work in the ACT-R cognitive
architecture which relies on the same
architectural mechanisms we will use in this model: production compilation and
the core-affect module \citep{larue2017core,larue2018cognitive}.
ACT-R 
is a theory of
human cognition and a computational cognitive architecture composed of
different modules (i.e., specialized processing units) that interact via their
associated buffers to complete a cognitive task \citep{anderson2013architecture}.
The distinction between two
systems, procedural memory (rules) and declarative memory (chunks), is a strong
foundational principle.  An estimate of \emph{human} real-time processing is
associated with rule-firing  (250 ms); similarly, a time estimate, a function
of recency and frequency of the use of that specific memory, is associated with
memory retrieval operations.

In ACT-R, the transition from novice to expert has been modeled as skill
learning through production compilation for specialization
\citep{taatgen2002children}. Production compilation is the learning mechanism
that supports the transition from novice to expert. Through production
compilation, two existing rules (from procedural memory) can be combined into
one new rule if they are fired sequentially frequently. This transforms
task-specific declarative knowledge and general production rules into
task-specific production rules. Chunk retrievals from declarative memory that
were occurring in each of the rules will be substituted by constants into the
new rule (one rule fires faster than two).

A model of memory athletics would be decomposed in four phases:
(1) Initialize the model,
(2) Build PAO of a deck's face and number cards,
(3) Rehearsal/Practice,
and (4) Competition.

During this learning phase, we emulate the knowledge a participant would have
prior to building a PAO and which will be necessary to build a PAO. In future
models, initial declarative memory chunks could be initialized with an open
database (ConceptNet \citep{speer2017conceptnet} + NRC emotion database \citep{mohammad2013crowdsourcing}), providing a semantic network of
chunks and NRC database for arousal and valence values of words.  Learning the
palace occurs through predictive learning. A palace is initialized as a
sequence of chunks (locations within the palace) from a preset map, and
situating (i.e. locating elements) in this memory palace will enable the
memorization of a sequence of elements during competition. Through multiple
simulations and learning by comparison with the ``real world,'' the palace
``geography'' will be reinforced and lead to production compilation (i.e., from a
procedure retrieving elements explicitly from declarative memory to procedures
including implicit/overlearned knowledge).

In the second phase, to build a PAO narrative of face cards of a deck
(following the procedure described by ND), the associate-face-cards rule
creates an association with the elements with the highest arousal. For example, when
a retrieval request will be made for the king, the PAO image with the highest
activation and arousal among those with some semantic connection to ``king''
(e.g., father) will be retrieved and a production rule will link the card to
the word. In S3Q terms, elements with the highest arousal are the one that
have been the most simulated and are the fastest to retrieve (maximizing the
feeling of rightness). Simulation in the ACT-R model is supported by
consecutive open-retrieval requests to declarative memory elements (i.e.,
chunks), retrieval requests with no constraints on chunk values but constraints
on activation parameters of arousal/core-affect value (i.e. to ensure the
retrieval of elements with a similar emotional intensity). To consolidate those
memory chunks, the model will go through a ``dreaming'' phase
\citep{juvina2018modeling}: a dreaming model will do ``open-retrievals'' (i.e.
retrieval requests to declarative memory with no constraints on specific value)
which will reinforce chunks with the highest activation (i.e. most recently
and/or frequently activated chunks). This mirrors the way in which the qualia
vocabulary underpinning the S3Q model is created and modified. The repetition
of this simulation and dreaming in this phase will lead to production
compilation of relationships in PAOs. Through production compilation, a model
instance might no longer need to retrieve PAO chunks from declarative memory. 

Following ND's verbal report of the task (i.e., face and number cards are
encoded in two separate steps), to encode other types of cards, a retrieve-else
rule will exclude retrieval of existing face card associations and loop to
associate any number-cards to a new ``person''. The retrieval request will
include characteristics of the card, and through pattern matching, the highest
arousal ``person'' chunk will be retrieved, and an object and action will be
retrieved in subsequent rules to be associated to the person in a final rule. 

During the third phase, rehearsal/practice, the model's instance will test the
strength of its associations for each card through free-recall rules. We
hypothesize that the differentiation between each type of action/object for
each person will occur naturally as ``testing'' the memory will result through
practice in the elimination of pairings which lead to confusion, maintaining
structural coherence. More specifically, structural coherence is gained by
eliminating interferences. The structural coherence of a simulation is
maximized when this simulation can not be confused with another one (i.e., in
this case structural coherence = more distinctiveness). Simulation will lead to
retaining the representation with the highest Feeling of Rightness (FOR) (i.e., the most consistent response modeled as the
most easily retrieved) and the least interference. This unique representation
will be a stable, consistent and useful representation. 

During the fourth phase, the competition, the model will leverage the mind
palace it has built to temporarily memorize new elements. The palace will
facilitate sequence learning: transition from room to room will have been
learnt through production compilation, and placing the characters which have
been rehearsed in every room rather than new characters will facilitate
retrieval: rehearsed characters and action are associated with a high feeling
of rightness (faster retrieval); placing them in different rooms of the
palace will allow for memorizing their sequence (the sequence of rooms is
present in implicit memory and will make for a faster use for encoding the
sequence of cards). We hypothesize that an accelerated retrieval is supported
by the compiled productions involved during competition, allowing for the
retrieval of association between characters/actions and rooms. 

\subsection{Future Work}
In this section we give two examples of planned experiments that would improve
our understanding of memory systems. The first shows how we can use the S3Q
perspective to make testable predictions about the impacts of modifications to
the Method of Loci strategy on memory capacity and failure rates. The second
tests the flexibility of the encoding scheme in order to tease out the complex
relationship between the neural mechanisms involved in the repetition-based
learning of PAO templates and the mapping to the phenomenal experience. 

\subsubsection{S3Q Flexibilities for Optimized Method of Loci}
The mainstream view is that consciousness consists of a small number of high
resolution items, specifying that this representation is limited by a number of
fixed-capacity pointers \citep{luck1997capacity}. Alternatively, an S3Q
two-system model requires a flexible resource allocation policy with a capacity
that is not fixed, but results from a complex interaction with complementary
nonconscious knowledge structuring \citep{mcclelland1995there}. For example,
there are separate pools of conscious resources for object and location process
codes in the brain, suggesting the possibility that PAO chunks and memory palace
representations flexibly use separate resources
\citep{baddeley1986,wickens1988codes}. Auditory perception resources are also
separable from visual, suggesting the use of auditory features to improve
encoding efficiency \citep{wickens2008multiple}. This modality could be beneficial for
distinctiveness, although this feature would ultimately compete for limited
conscious resources. \citet{alvarez2004capacity} found that estimated working
memory capacity of research participants in basic laboratory paradigms
decreased as a function of the difficulty of discriminating different items,
with maximally discriminable stimuli having a maximum capacity of approximately
4.5 for the simplest items. It is expected that a 4-item mnemonic system might
improve Speed Cards performance, and leveraging the auditory modality could
help this system push the limits of memory. 

S3Q theory suggests the generated
memory palace mnemonic system must suitably balance the number of items being
processed with the relationships between them during the Speed Cards
competition as more items per chunk will come at the cost of distinctiveness in
representation, and therefore memory. With practice, it is expected this can
become ultimately distinct. For example, a Person-Action-Object system is
expected to be a slower strategy for encoding a deck of cards, yet more
accurate than a Person-Action-Object-Sound system\footnote{
There may be additional limitations that arise here in finding 52 distinctive sounds to make this proposed method tractable.},
which would be predicted to
have the capacity to encode more cards per working memory cycle, but at the
risk of increased memory failures. Targeted Memory Reactivation (TMR) offers a
cutting-edge technique to test this \citep{rudoy2009strengthening}. It has been
demonstrated that pairing a tone with a learning event, and then replaying that
tone during sleep improves memory \citep{rudoy2009strengthening}. When these
tones are cued during slow wave sleep, techniques such as Representation
Similarity Analysis \citep{kriegeskorte2008representational} can reveal that
pattern of electrical activity in the neural structure that was seen during
learning is reactivated during spindle events between the hippocampus and
neocortex. Memories are actively labile during sleep and can be actively
manipulated as they are malleable. This technique could be employed in
competitive memory training research protocols to accelerate practice as well
as test the hypothesis that these different modalities are separate resources
in the single bound conscious representation. 

\subsubsection{Implicit Learning is Rapid but Inflexible}
We hypothesize that experts demonstrate operating characteristics of
representations with matured implicit circuits, such as affording rapid
response at the cost of flexibility. Thus, we propose that the physical card is
automatically mapped to an image of the person, action, or object in the memory
palace space of experts (i.e. card image $\to$ C3P0 image), while in novices there
are intermediate steps (e.g., card image $\to$ verbal ``three of clubs'' $\to$ C3P0
image). To test this, athletes can be presented with a series of deck
memorization tasks, in which experimenters vary the deck image to disrupt the
hypothesized implicit mapping from physical card image to simulated PAO image.
For example, variations of card images can be used: traditional English-style
cards, cards in which the suit colors are replaced (e.g. blue diamonds), or
cards in which the name of the card is written in English text, e.g. ``three of
clubs.''  Should the direct visual mapping exist, we predict that experts will
rapidly memorize the English-style cards, yet be slower for other deck images
\footnote{These predictions are in-line with ND's reported experiences using
similar decks (e.g., a ``JUMBO'' Bicycle deck rather than a standard deck) which
does not impact memorization times versus decks with more distinctiveness (e.g., a
french deck with the letters ``V,'' ``D,'' and ``R'' in place of the ``J,'' ``Q,'' and
``K'' in an english deck), where there is a noticeable hesitation.}
due to the distance in the perceptual space of situations.  On the other hand,
if experts are mapping to the card name, the verbally descriptive cards should
yield the fastest times, as would be predicted in novices -- which offers a
within-subjects design to see how this flexibility and expertise change as a
function of practice repetitions.

\section{Conclusions}
There are two types of learning and memory, and they interact to produce expert
performance. One type utilizes relatively fixed circuits honed for rapid
application of expertise, but rendered inflexible in transfer to novel
conditions. The other type balances S3Q to achieve flexible one shot learning.
The capacity of this second system is not fixed, but results from a balance of
S3Q resources interacting with the neuroplasticity mechanics of parallel
implicit representations. The theoretical position taken here aligns with the
assertion that working memory mechanisms affording one-shot learning are thus
not of a fixed capacity, but are optimized in interaction with implicit
mechanisms through practice. While memory athletes are pushing the limits of
the method of loci through high repetitions training of chunked information, it
will be important for generated mnemonic systems to balance the number of items
being represented with the distinctiveness of the features based on the limited
conscious resources of the brain. State-of-the-art memory techniques, such as
TMR, offer novel approaches to testing these theories while optimizing memory
performance for expert competition. The important future questions still
remain: in what ways are these multiple memory systems representations similar
and dissimilar, and what are the affordances each provide to both biological
and artificially intelligent implementations.

\appendix
\renewcommand\thesection{Appendix \Alph{section}}
\section{The Dominic System}
\label{appendix:dominic}

The overall approach of the Dominic System is to assign visuals, also known as
``attributes,'' to each card. This system assigns each of the 40 non-face cards a
unique pair of letters where the first letter represents the number (e.g., 1 =
A, 2 = B, 3 = C, ...) and the other letter represents the suit (H = Hearts, D =
Diamonds, C = Clubs, S = Spades). (Traditionally, no mapping is given for the
twelve face cards since these are usually easier to visualize due to the
intrinsic semantics as described below.) Next, a person (or animal, fictional
character, etc.) who has those initials is chosen. For example, the 5 of Clubs
is assigned the letter pair (E, C), and ND chose Eric Clapton. The face cards
are handled differently. They are chosen to be intuitive based on the suit and
face. For example, consider the King of Hearts. The heart suit reminds ND of
family and the ``king'' of his family is his father, thus ND choses his father
for the King of Hearts. This gives a card $\to$ person mapping. It is important
that this mapping associates a distinctive person with each card. That is, once
a person is associated with one card, that person (or subjectively similar
people) should not be associated with another card. To memorize a deck of cards
with this mapping, memory athletes would need a memory palace with 52 positions. The
first card is mapped to its person and a mental image of the person is stored
in the first position of the memory palace. This repeats for the whole deck of
cards. Once the deck of cards is stored in the memory palace, memory athletes
recall the deck by traversing the memory palace and reversing the mapping at
each location to recall the card in that position.

This limited system has the disadvantage of requiring 52 positions in the
memory palace, but the full PAO system only requires 17 (or optionally 18). The
system starts with the person mappings assigned to each card. Then, an action
and an object are assigned to each card that corresponds naturally with the
person already assigned to that card. The action and object are chosen to
correspond with the person to make learning the mapping easier. For an example
of an action and object, recall that the person assigned to the 5 of Clubs is
\PAO{Eric Clapton}. The action is chosen to be \PAO{playing} an instrument, and the object is a
\PAO{guitar}. It is important that the action and object are both unique. That
is, no action or object may be assigned to two different cards. At this point,
a card $\to$ (person, action, object), or PAO, mapping has been constructed for all
52 cards.\footnote{A complete PAO list for a 52 card deck used by ND can be
found in his recent book \citep{dellis2018remember}.}

To memorize a deck of cards with this mapping, memory athletes use a memory
palace with 17 (or optionally 18) positions. The first card is mapped to the
first card's person, the second card is mapped to the second card's action, and
the third card is mapped to the third card's object. A mental image is formed
of the person associated with the first card doing the action associated with
the second card to the object associated with the third card. (For instance, 
\card{3C}\card{5C}\card{7H} would be \PAO{C-3PO} \PAO{playing} a \PAO{pizza} as a guitar.) That
image is then placed in the first location of the memory palace, which has been
previously constructed. Then the fourth, fifth, and sixth cards are likewise
mapped to an image and that image is stored in the second position of the
palace. This repeats for the first 51 cards/17 positions. The final card is
remembered without the memory palace and, if forgotten, deduced by process of
elimination (or stored in the 18th position). To recall the deck of cards,
memory athletes traverse the memory palace. In order to construct the memory
palace itself, the memory athlete chooses a well-known real world place (ND
uses places such as his home or high school campus), and rehearses a specific
set of locations in those places which will serve as the locations at which
visual items will be stored. For example, a memory palace through a high school
might start at the entrance to the school, then proceed to a hallway just
inside the door, and then to a classroom nearby with the distinct locations
connected by a rehearsed path through the palace.

Memory athletes do not need to follow the system for choosing people exactly as
described. To enable faster learning, it is often better to choose something
intuitive for number cards, as is done for face cards, so that the mapping is
easier to remember. In this way, less time is spent thinking of who a card maps
to. For example, ND struggled to come up with a sufficiently intuitive mapping
for the 4 of Spades. The 4 of Spades maps to (D, S). This reminded ND of the
Nintendo DS and, in turn, Super Mario. So, the person for the 4 of Spades is
Mario.

In order to make the mappings intuitive, the PAO associations may evolve over time. Early on, some
associations may be found to be weak (e.g., hard to visualize or hard to recall). These can be
replaced. Likewise, memory athletes may find new things they find memorable that they want to
include in their PAO mappings. Eventually, mapping a card to its associations and vice versa becomes
automatic. Memory athletes look at a card and automatically think of the person associated with it.
For Speed Cards, this is critical because the speed of memorization often comes down to how quickly
memory athletes can encode the cards.

It should be noted that, while the technique has been described as primarily visual, it does not
need to be. Memory athletes may also leverage auditory imagination, somatosensory imagination, etc.
ND reports using non-visual imagery in his novice stage to make the mental image as rich as
possible. For example, ND would try to hear particular accents. At the expert level, ND reports
using the sound of Arnold Schwarzenegger's voice as the main representation for Schwarzenegger as a
person in his scheme. ND reports that using sound as the main representation for a person is
uncommon in his encodings and Schwarzenegger may be unique in this respect. And, ND reports using
auditory imagery for some actions, like shooting.

\bibliographystyle{plainnat}
\bibliography{references}

\end{document}